\documentclass[10pt,a4paper,twoside]{article}
\usepackage{epsfig}
\usepackage{baltlat6}
\usepackage{array}
\begin{document}
\ \
\vspace{0.5mm}
\setcounter{page}{277}
\vspace{8mm}

\titlehead{Baltic Astronomy, special volume of Serbian conference...}

\titleb{OPTICAL SPECTROSCOPY WITH THE TECHNOLOGY\\
 OF VIRTUAL OBSERVATORY}

\begin{authorl}
\authorb{P.~\v{S}koda}{}
\end{authorl}

\begin{addressl}
\addressb{}{Astronomical Institute of the Academy of Sciences,
 Fri\v{c}ova 298, 251\,65\ Ond\v{r}ejov, Czech Republic}
\end{addressl}

\submitb{Received: 2011 June 10; accepted: 2011 December 15}

\begin{summary} 
The  contemporary astronomy is flooded with an exponentially growing
petabyte-scaled data volumes produced  by powerful ground and space-based
instrumentation as well as a product of extensive computer simulations and
computations of complex numerical models. The efficient organisation and
seamless handling of this information avalanche stored in a  world-wide spread
heterogeneous databases and the facilitation of extraction of new physical
knowledge about the Universe is a primary goal of  the rapidly evolving
astronomical Virtual Observatory (VO). We give an overview of current
spectroscopic capabilities of VO and identify the future
requirements indispensable for detailed multi-wavelength analysis of huge
amounts of spectra in a semi-automatic manner.
\end{summary}


%% \resthead is the RUNNING TITLE at top of the pages
\resthead{Optical spectroscopy with the technology of Virtual Observatory}%
{P.~\v{S}koda}

%-------------------------------------------------------------------

\sectionb{1}{DATA AVALANCHE IN ASTRONOMY} 

The modern instrumentation of large telescopes like large mosaics of CCD chips,
massive multi-object spectrographs with thousands of optical fibers or
microlenses in Integral Field Units (IFU) as well as fast radio correlators
mixing inputs of tens of antennas have been producing terabytes of raw data per
night and for their reduction a grids of supercomputers are needed.  For
example the future all sky survey LSST will yield 30TB od raw data every night
requiring for the reduction of data the processing power about 400 TFLOPs%
\footnote{\tt http://www.lsst.org/lsst/science/technology} (Ivezi\'c et al. 2011).

The current growth of astronomical data in large archives has been rising
exponentially with the doubling constant less than 6--9 months. It is much
steeper that the famous Moore's law of technology advances which predicts the
doubling time of computer resources about 18 month (Quinn et al. 2004, Quinn 2007).

Astronomy is just facing the avalanche of data that no-one
can process and exploit in full.  It is clear that such data cannot be
processed and analysed in a classical manner on local
desktop and the concept of remote processing has to be introduced.

The promising solution of handling this data deluge is the implementation of
service oriented architecture moving the burden of data processing,
pre-analysis and searching towards the high performance well equipped data
centres.

\sectionb{2}{VIRTUAL OBSERVATORY} 

The key role in this effort plays the concept of Virtual Observatory, whose
goal is to provide standards describing all astronomical resources worldwide
and to enable the standardized discovery and access to these collections as
well as powerful tools for scientific  analysis and visualisation% 
\footnote{http://www.ivoa.net/pub/info/TheIVOA.pdf}.

As the  VO mostly provides access to final science-ready data, the VO data
provider has to make the final calibrated data VO-compatible. This requires
creation of a set of metadata (for curation, provenance and characterization)
and preparation of access interface in accordance with appropriate VO standard
protocols%
\footnote{http://www.ivoa.net/Documents/Notes/IVOAArchitecture/index.html}.

Most of the highly acknowledged astronomical services like Vizier, Simbad, NED
or tools as Aladin are the practical examples of VO technology  in everyday
use.  All the complexity of replicated database engines, XML processors, data
retrieval protocols as well as distributed grids of supercomputers providing
powerful services is hidden under the  hood of a simple web-based form
delivering complex tables, images, previews, graphs just on the button click.

\subsectionb{2.1}{VO and Astronomical Community}

The VO development and preparation of standards is coordinated by the
International Virtual Observatory Alliance (IVOA) currently having 17 national
and 2 multinational (ESA and ESO) members% 
\footnote{http://www.ivoa.net/pub/info}.
The International Astronomical Union (IAU) is promoting the VO through its
Commission~5 and dedicated VO working group. 

The key task of IVOA is the design of global standards (data models, data formats,
protocols) for the whole VO infrastructure reflecting the needs and priorities
of different  astronomical communities and projects.  There is also an Astronet
strategic plan for European astronomy created by several funding agencies to
establish a comprehensive long-term planning for the development of European
astronomy -- The Infrastructure Roadmap% 
\footnote{http://www.astronet-eu.org/IMG/pdf/Astronet-Book.pdf}. 
It emphasises the role
of VO for future astronomers and requires all new projects data to be
VO-compliant. 

\subsectionb{2.2}{Interoperability} 

Although very sophisticated most of current archives are just isolated islands
of information with unique structure, data formats and access rules (including
specific search engine). Even if the questions asked are simple the returned
data have different scales, orientation, astrometric accuracy as well as
coordinate system.

Thus the key issue for success of interoperability of distinct services is the
strict standardization of data format of data content. The astronomy
has an advantage of using the same format --- FITS --- of all astronomical frames for decades.
The global interoperability of VO infrastructure is based on
several standardized components:

\subsubsectionb{2.2.1}{VOTable}

The bunch of data (e.g. columns of numbers) are not exploitable without
metadata (i.e. labels of columns and README explaining the labels).  The
metadata are describing the same physical variables with the same term despite
the original label used in the given table. The same with units. The important role
play here the controlled semantic vocabulary of Unified Content Descriptors
(UCD)% 
\footnote{http://www.ivoa.net/Documents/latest/UCDlist.html}.

This together with the standardized access protocols allow to design clients
which can query and retrieve data from all VO-compatible servers at once.
Standard data format in VO, the VOTable%
\footnote{http://www.ivoa.net/Documents/VOTable}, is a XML standard allowing
full serialization (first are sent metadata and than follows a stream of
numbers) and embedded hyperlinks for real data contents (e.g. URL to FITS on
remote servers).

All the available astronomical knowledge about acquisition process,
observing conditions as well as the whole processing and reduction  should be
included in the self-describing part of VOTable metadata (called provenance)
together with all proper credits and citations (called curation metadata)%
\footnote{http://www.ivoa.net/Documents/latest/RM.html}.

All the physical properties of observation should be placed in other part of
meta-data called characterisation which should describe all the relevant
information about spatial, temporal and spectral coverage, resolution,
position, exposure length, filters etc.%
\footnote{http://www.ivoa.net/Documents/latest/CharacterisationDM.html}

\subsubsectionb{2.2.2}{VO Registry}

The worldwide knowledge about the particular VO resource requires the global
distributed database similar to Internet Domain Name Service (DNS).
So all VO resources (catalogues, archives, services) have to be registered in
one of the VO Registries%
\footnote{http://www.ivoa.net/Documents/RegistryInterface}.
The registry records are encoded in XML.
Every VO resource has the unique identifier looking like URL but instead of
\verb+http://+ having the prefix  \verb+ivo://+\footnote{http://www.ivoa.net/Documents/latest/IDs.html},
which is considered to be compulsory for referring to datasets in some journals. 

All the information describing the nature of the data, parameters,
characterization or even references and credits put in one registration server
are being regularly harvested by all VO registries, so every desktop client may
have the fresh list of everything  available in VO.

\subsubsectionb{2.2.3}{Data Access Protocols}

The transparent access of data from VO servers is accomplished using a number
of strictly controlled  protocols. Among the most commonly used belong:
\begin{description} 
\item[ConeSearch] It returns the catalogue information about objects in given
circle (position, radius) on the celestial sphere%
\footnote{http://www.ivoa.net/Documents/latest/ConeSearch.html}.
\item[SIAP] The Simple Image Access Protocol is intended for transfer of images
or their part of given size and orientation%
\footnote{http://www.ivoa.net/Documents/latest/SIA.html}.
\item[SSAP] The Simple Spectra Access Protocol is designed to retrieve spectrum
of given properties (time, position, spectral range, spectral resolution power
etc.)%
\footnote{http://www.ivoa.net/Documents/latest/SSA.html}.
\item[SLAP] The Simple Line Access Protocol, mostly used in theoretical
services returns the atomic or molecular data about given line transitions in
selected wavelength or energy range and vice versa%
\footnote{http://www.ivoa.net/Documents/latest/SLAP.html}.
\item[TAP]  The Table Access Protocol%
\footnote{http://www.ivoa.net/Documents/TAP} is a complex protocol for querying
very large  tables (like catalogues, observing logs etc.) from many distributed
servers simultaneously (it has asynchronous mode for very long time queries
based on Universal Worker Service Pattern (UWS)%
\footnote{http://www.ivoa.net/Documents/UWS}).  
\end{description}

The queries are written using  the specific superset of SQL, called ADQL%
\footnote{http://www.ivoa.net/Documents/latest/ADQL.html}
(Astronomical Data Query Language)  with operators allowing selection of
objects in sub-region of any geometrical shape on the sky or the XMATCH operator
allowing to decide the probability of match of two sets of objects in two
catalogues with different error box (called cross-matching of catalogues)%
\footnote{http://voera.ncsa.uiuc.edu/course/adql.pdf}.

\subsubsectionb{2.2.4}{VO Applications}

The interaction of VO infrastructure with end user (scientist) is provided by a
number of VO-compatible applications.  Most of them are desktop clients (written
in the multi-platform manner --- in Java or Python) There are general
tools for work with multidimensional data sets --- 
VOPlot\footnote{http://vo.iucaa.ernet.in/voi/voplot.htm} or
TOPCAT\footnote{http://www.star.bris.ac.uk/~mbt/topcat/},
celestial atlases for showing images over-plotted with catalogue data --- Aladin%
\footnote{http://aladin.u-strasbg.fr/aladin.gml}
or VIRGO\footnote{http://archive.eso.org/cms/tools-documentation/visual-archive-browser}, as well as applications for specific operations on spectra ---
SPLAT\footnote{http://star-www.dur.ac.uk/~pdraper/splat/splat-vo},
VOSpec\footnote{http://www.sciops.esa.int/index.php?project=ESAVO\&page=vospec} and 
SpecView\footnote{http://www.stsci.edu/resources/software\_hardware/specview}. 
The regularly updated list of all VO applications is maintained at EURO-VO Software 
page\footnote{http://www.euro-vo.org/pub/fc/software.html}.

The so far tedious but very important astronomical technique is the
determination of spectral energy distribution (SED), which helps to reveal the
physical nature of the astronomical object. The VO technology can help a lot in
an aggregation  of observed data and theoretical models.  Building of SEDs in VO
consists of collecting the scattered photometric data, their transformation
into common filter system (using the database of different filter transmission
curves) and fitting theoretical model obtained as well from VO databases of
model spectra.  

One recent application for building SEDs is VAO Iris%
\footnote{http://cxc.harvard.edu/csc/temp/sed/intro/index.html} incorporating
the advanced package for fitting spectra Sherpa%
\footnote{http://cxc.cfa.harvard.edu/sherpa4.3/index.html}.
Some more complicated tools are being built as  web services or web
applications (with query forms etc.). 
The example of very usefull web-based
application is Virtual Observatory SED Analyzer  (VOSA)%
\footnote{http://www.laeff.inta.es/svo/theory/vosa2}.

As every application is written by different developers having in mind specific
type of scientific analysis, there does not exist any single complex
all-purpose VO tool. Instead of this, in the spirit of UNIX thinking, the
isolated applications have common interoperability interface using the Simple
Application Messaging Protocol (SAMP)%
\footnote{http://www.ivoa.net/Documents/SAMP/index.html}.
VO applications supporting SAMP can exchange their data (VOTables, spectra,
images) with other SAMP-compatible application. This allows (together with
command scripting) the building of complex processing and analysing workflows
by chaining the VO applications. 

\subsectionb{2.3}{Science with VO}

The key advantage of maintaining VO infrastructure is the new type of science
called VO-science.  The huge data-mining potential and multi-wavelength nature
of VO infrastructure allows to tackle problems not feasible by any other means (e.g.
search of rare events, classes of objects, pan-spectral research from gamma to
radio etc.)
There is  already number of referred articles using VO in astronomical
research.  Current list of VO-based papers is maintained at EURO-VO web%
\footnote{http://www.euro-vo.org/pub/fc/papers.html}.

An example of VO power is the study of rare objects.  In the 2005 VO Science
demonstration were found 100 new candidate in transition phase from AGB star to
planetary nebulae using VO methodology in addition to 200 already known so
far (Tsalmantza et~al. 2006).
Another example is the discovery of a new brown dwarfs  
 by using VO tools to cross-match two large catalogues%
\footnote{http://www.euro-vo.org/pub/fc/workflows/BDs.html}.
VO enabled the discovery of extremely bright white subdwarf
(Caballero \& Solano 2007) and many other extreme objects.

Success story of using complex VO technology (spectral fitting,  photometric
search, catalogues cross-matching)  and data mining technology to yield new
scientific results not achievable by any classic method (due to its huge
scale) were justified  by  Chilingarian et~al. (2009).

\subsectionb{2.4}{Theory VO}

Not only observational astronomy is producing large data volumes. The same data
format and protocols are used for access to theoretical spectra or artificial
images of simulated stellar clusters, models of stellar atmospheres, 
isochrones in stellar evolution models or to results of  simulations of galaxy
collisions or even evolution of all Universe.

The  metadata and query parameters are model
specific (e.g. $T_{\rm eff}$, $\log g$), but the result is output as VOTable.
Special data and metadata access protocols have been used and new are still
suggested in the IVOA. The specific data model%
\footnote{http://www.ivoa.net/cgi-bin/twiki/bin/view/ivoa/ivoatheorysimdmspec}
and 
database structures called SimDB%
\footnote{http://www.ivoa.net/cgi-bin/twiki/bin/view/ivoa/ivoatheorysimdb}
are used as the backbone of large simulation projects as is the Interstellar Media
Platform%
\footnote{http://www.ivoa.net/internal/IVOA/InterOpMay2010Theory/IVOA10\_Victoria\_PDR.pdf},
 part of which is the code for computation of physical parameters in
Photo-Dissociation Regions (PDR)%
\footnote{http://pdr.obspm.fr/PDRcode.html}.
The large EU-FP7 project Virtual Atomic and
Molecular Data Centre%
\footnote{http://www.vamdc.org}
will use the VO  infrastructure for accessing distributed
atomic and molecular databases as well as SLAP protocol for client access.

%===================================================================
%===================================================================

\sectionb{3}{OPTICAL SPECTROSCOPY WITH LARGE DATA SETS}

There is a number of spectroscopic techniques requiring processing of large
amount of spectra of the similar size and resolution to get the physical
information about single object.  We try to identify the ones, where the VO
technology could help to reduce the tedious work and speed up the processing.
More detailed description of various spectroscopic techniques in VO is given in 
\v Skoda (2008) and \v Skoda (2009a,b).

Accomplishing the multi-spectral analysis in VO environment may benefit from
automatic aggregation of distributed archive resources, seamless on-the-fly
data conversion, common interoperability of all tools  and powerful graphical
visualisation of measured and derived quantities.

\subsectionb{3.1}{Simple Visualisation of Spectra Changes}

A lot of the information about the behaviour of astronomical objects can be
estimated just by a visual inspection of a spectrum (spectral type,
peculiarity, emission) or a time series of spectra (pulsations, binarity).  The
basic method is the over-plotting of many spectra in the same units and scale.
It may be very efficient with VO-enabled tools obtaining the number of spectra
cached  immediately from VO spectral servers  with SSAP.  
There are many possibilities of plots. Among  the most  common belong:

\subsubsectionb{3.1.1}{Stacked Line Profiles} 

The high resolution spectra with high SNR may reveal on some objects small
variations of the   profile of spectral lines.  The study of LPV requires many
(even hundreds) of spectra to be over-plotted with the additional vertical offset
(corresponding to time of observation or just a convenient constant) to see
the changes easily.  Sometimes the animation of changes in individual spectral
line is very impressive.  Such a plot is helpful in asteroseismology or for
estimating changes in stellar  winds.

\subsubsectionb{3.1.2}{Dynamic spectrum} 

It is sometimes called the grey representation or trailed spectrum. The basic
idea is to find the small time-dependent deviations  of individual line
profiles from some average.  The recipe is simple.  First the average of many
high dispersion high SNR spectra (with removal of outliers) is prepared (called
template spectrum). Then each individual spectrum in time series is either
divided by the template (quotient spectrum) or the template is subtracted from
it (the differential spectrum). The group of similar resulting intensities is
given the same colour or  level of gray.  Examples may be found in 
de~Jong~et~al. (1999) or Maintz (2003).

\subsectionb{3.2}{Complex Processing Methods }

For the complex techniques given below, a lot of additional information is
required in addition to spectral data.  Programs require complicated
configuration files in given format and some interactive trials to find the
best results using the output from recent run as input to the next one. They are
written often in FORTRAN without graphical interface and often without the plotting
capabilities.  They are designed for batch runs driven by  parameter files.
The VO tools can help to collect (aggregate) the required spectra and preselect
them (in order to remove bad quality data, select given time or spectral range
or to isolate interesting spectral features) before entering the specific processing,
but they can help in accessing the databases of theoretical models as well.

\subsubsectionb{3.2.1}{Doppler Imaging }

It was discovered by Vogt \& Penrod (1983a) as a method allowing the surface
mapping of stellar spots.  First test were done on stars of RS CVn type and on
$\zeta$~Oph (Vogt \& Penrod 1983b). The method works well on rapid rotators and needs a
high resolution spectra with very high SNR (300--500). Tho whole rotational
period should be covered well, better several times.  When all the requirements
are met,  the map  of surface features (spots, nodes of non radial pulsations)
is obtained with very high accuracy.  A recent application of this technique on
$\zeta$~And is given by Korhonen et~al. (2010).

\subsubsectionb{3.2.2}{Doppler tomography}

It was introduced by Marsh \& Horne (1988) for mapping the distribution of
emitting circumstellar matter in binary system. One of the successful applications
gave a picture of   accretion jets in Algols (Richards 2004).  It uses trailed
spectrum  in velocity scale. The result is 2D image in velocity space.  The
transformation of radial velocity space to coordinate space is ambiguous, which
causes problems in interpretation of Doppler tomograms. 

\subsubsectionb{3.2.3}{Zeeman Doppler Imaging}

Quite complicated processing of spectra is required for study of stellar
magnetic fields.  The estimation of magnetic field from  polarimetry using the
Zeeman effect involves the processing of long series of homogeneous spectra
to be accomplished in parallel with extreme precision and requires  the
information from synthetic models (simulation of Stokes parameters on simulated
magnetic stars) The nice example is the model of II Peg by 
Carroll et~al. (2007).

\subsubsectionb{3.2.4}{Spectra Disentangling}

This method allows to separate the spectra of individual stars in binary or
multiple systems and simultaneously to find orbital parameters of the system,
even in case of heavy blending of lines.  It supposes the changes in line
profile are caused mainly by combination of Doppler shifted components. The best
solution of orbital parameters and disentangled line profiles of individual
stellar components are found by least square global minimisation.  The method
also enables to  remove the telluric lines with great precision.  
Although there are several methods of spectra disentangling, the most
commonly used is the Fourier space disentangling introduced by Hadrava(1995) 
in program KOREL with several generalizations of the methodology. 
The web service VO-KOREL
\footnote{http://stelweb.asu.cas.cz/vo-korel}
is based on VO technology of Universal Worker Service
({\v S}koda \& Hadrava 2010).

\sectionb{4}{ASTROINFORMATICS}

As was said above, the current science is commonly understood to be
data-intensive or data-driven.  The  research in almost all natural sciences is
facing the 'data avalanche' represented  by exponential growth of information.
The effective retrieval of a scientific knowledge from petabyte-scale databases
requires  the qualitatively new kind of scientific discipline called e-Science,
allowing the global collaboration of virtual communities sharing the enormous
resources and power of supercomputing grids (Zhao et~al. 2008
and Zhang et~al. 2008).
E-Science is often referred to as  the internet-enabled sharing of distributed
data, information, computational resources, and team knowledge for the
advancement of science.  As an example of working e-Science technology in
astronomy is given the emerging new kind of astronomical research methodology
--- the Astroinformatics.

It is based on systematic application of modern informatics and advanced
statistics on huge astronomical data sets. Such an approach, involving the
machine learning, classification, clustering and data mining yields the new
discoveries and better understanding of nature of astronomical objects.  The
Astroinformatics is an example of a new science methodology  where the new
discoveries result often from the searching of outliers in common statistical
patterns.  It is sometimes presented as new way of doing
astronomy (Borne et~al. 2009, Ball and Brunner 2010).
Examples of successful application of astroinformatics in spectroscopy is the
data mining of spectra with lines of given shape (V\'a\v{z}n\'y 2011) and the
estimation of photometric red shifts (D'Abrusco et al. 2009).

\sectionb{5}{CONCLUSIONS}

Astronomical spectroscopy uses a wide range of techniques with different
level of complexity to achieve its final goal --- to  estimate the most
precise and reliable  information about celestial objects. The  large part
of spectroscopic analysis today  has been accomplished by several
independent non VO-compatible legacy packages, where  each works  with
different local files in its own data format. Analysis of large number of
spectra is thus very tedious work requiring good data bookkeeping. 

Accomplishing the analysis in VO infrastructure may benefit from automatic
aggregation of distributed archive resources (e.g. the multispectral
research), seamless on-the-fly data conversion, common interoperability of
all tools and powerful graphical visualisation of
measured and derived quantities. 

Combining the VO infrastructure power and the easy and transparent high
performance computing on GRID will allow the advanced analysis of  large
spectral surveys  feasible in a reasonable time.  The crucial role in
understanding the results of such an analysis plays the Astroinformatics as a
methodology allowing the extraction of new physical knowledge from astronomical
observations, what is the final goal of all scientific effort.

\thanks{This work has been supported by grant III 44002 of
Ministry of Education and Science of Republic of Serbia and EURO-VO ICE project.
 The Astronomical Institute Ond\v{r}ejov is supported by project AV0Z10030501}

\References

\refb Ball, N.~M., and Brunner, R.~M. 2010, International, Journal of Modern
Physics D 19, 1049 (arXiv:0906.2173v2)

\refb Borne, K., et~al. 2009, in {\it Astro2010: The Astronomy and Astrophysics
Decadal Survey, Position Papers}, 6 (arXiv:0909.3892)

\refb Caballero, J.~A. \& Solano, E. 2007, ApJ. Lett., 665, L151

\refb Carroll, T.~A., Kopf et~al. 2007, Astronomische Nachrichten, 328, 1043 

\refb Chilingarian, I. et-al. 2009, Science, 326, 1379

\refb D'Abrusco, R., Longo, G., and Walton, N.A. 2009, MNRAS, 396, 223 (arXiv:0805.0156)

\refb  de~Jong, J.~A. et~al.,1999 A\&A, 345, 172 

\refb Hadrava, P.\ 1995, A\&AS, 114, 393

\refb Ivezi\'c, \v{Z}. et al. 2011, {\it LSST: From Science Drivers to Reference Design and Anticipated Data Products},
arXiv:0805.2366

\refb Korhonen, H. et~al.\ 2010, A\&A, 515, A14 (arXiv:1002.4201)

\refb Maintz, M. 2003, {\it Be binary stars with hot, compact companions}, PhD.
thesis, University of Heidelberg

\refb Marsh, T.~R., \& Horne, K.\ 1988, MNRAS, 235, 269 

\refb Quinn, P.  2007, {\it Data Intensive Science needs for Australian
Astronomy}\\
{\small\tt http://astronomyaustralia.org.au/ASTRO-projects-infrastructure.pdf}

\refb Quinn, P., Lawrence, A., Hanisch, R. 2004, {\it The Management, Storage
and Utilization of Astronomical Data in the 21st Century}\\
{\small\tt http://www.ivoa.net/pub/info/OECD-QLH-Final.pdf}

\refb Richards, M.~T.\ 2004, Astronomische Nachrichten, 325, 229 

\refb {\v S}koda, P.\ 2008, in {\it Astronomical Spectroscopy and Virtual
Observatory}, eds. M.~Guainazzi \& P.~Osuna, ESA, 97 

\refb {\v S}koda, P.\ 2009a, Mem. Soc. Astron. Ital., 80, 484 

\refb {\v S}koda, P.\ 2009b, in {\it Multi-wavelength Astronomy and Virtual
Observatory}, eds. D.~Baines \& P.~Osuna, ESA, 11 

\refb {\v S}koda, P., \& Hadrava, P.\ 2010, in {\it Binaries – Key to
Comprehension of the Universe}, eds. A.~Pr\v{s}a and M.~Zejda,  ASP Conf. Ser.,
435, 71 (arXiv:1003.4801)

\refb Tsalmantza, P. et~al. 2006, A\&A, 447, 89

\refb V\'a\v{z}n\'y, J. 2011, {\it Virtual Observatory and Data Mining}, Master thesis, Masaryk University, Brno
http://is.muni.cz/th/211665/prif\_m/thesis.pdf

\refb Vogt, S.~S., \& Penrod, G.~D.\ 1983a, PASP, 95, 565 

\refb Vogt, S.~S., \& Penrod, G.~D.\ 1983b, ApJ, 275, 661

\refb Zhang, Y., Zheng, H., \& Zhao, Y. 2008, in {\it SPIE Conference Proceedings},
7019, 108

\refb Zhao Y., Raicu, I., \& Foster, I. 2008, in {\it IEEE Congress on Services
Part I}, 467--471 (arXiv:0808.3545)

\end{document}